\documentclass[runningheads]{llncs}
\usepackage{graphicx}
\usepackage{rotating}
\usepackage[]{mdframed}

\usepackage{color}

\usepackage[bookmarks, colorlinks, breaklinks]{hyperref}
\usepackage{sidecap}

\newcommand{\cC}{\mathcal{C}}
\newcommand{\cG}{\mathcal{G}}
\newcommand{\beq}{\begin{equation}}
\newcommand{\eeq}{\end{equation}}
\newcommand\blinded[1]{#1}
\newcommand\noshow[1]{}

\begin{document}
\title{
 Overstatement-Net-Equivalent Risk-Limiting Audit: ONEAudit}
  \titlerunning{ONEAudit}

\author{
\blinded{Philip B. Stark\inst{1}\orcidID{0000-0002-3771-9604} }
}
\authorrunning{\blinded{P.B. Stark}}
%
\institute{\blinded{University of California, Berkeley, CA USA}
\email{\blinded{stark@stat.berkeley.edu}}}
\maketitle              
\begin{abstract}
    A procedure is a risk-limiting audit (RLA) with risk limit $\alpha$ if it has probability at least $1-\alpha$ of correcting
    each wrong reported outcome and never alters correct outcomes.
    One efficient RLA method, card-level comparison (CLCA), compares human interpretation 
    of individual ballot cards randomly selected from a trustworthy paper trail to the voting system's interpretation 
    of the same cards (cast vote records, CVRs).
    CLCAs heretofore required a CVR for each cast card and a ``link'' identifying which CVR is for which card---which
    many voting systems cannot provide.
    This paper shows that every set of CVRs that produces the same aggregate results overstates contest margins by the same amount:
    they are \emph{overstatement-net-equivalent} (ONE).
    CLCA can therefore use CVRs from the voting system for any number of cards and ONE CVRs created \emph{ad lib}
    for the rest.
    In particular:
    \begin{itemize}
    \item Ballot-polling RLA is algebraically equivalent to CLCA using ONE CVRs derived from the overall contest results.
    \item CLCA can be based on batch-level results (e.g., precinct subtotals) by constructing ONE CVRs for each batch.
    In contrast to batch-level comparison auditing (BLCA), this avoids manually tabulating entire batches and 
    works even when reporting batches do not correspond to physically identifiable batches of cards, when BLCA is impractical.
    \item If the voting system can export linked CVRs for only some ballot cards, auditors can still use CLCA
    by constructing ONE CVRs for the rest of the cards from contest results or batch subtotals.
    \end{itemize}
    This works for every social choice function for which there is a known 
    RLA method, including IRV.
    Sample sizes for BPA and CLCA using ONE CVRs based on contest totals are comparable. 
    With ONE CVRs from batch subtotals, sample sizes are smaller than than for BPA when batches are 
    homogeneous, approaching those of CLCA using CVRs from the voting system,
    and much smaller than for BLCA:
    A CLCA of the 2022 presidential election in California at risk limit 5\% using ONE CVRs for 
    precinct-level results would sample approximately 70~ballots statewide, if the reported results are accurate,
    compared to about 26,700 for BLCA.
    The 2022 Georgia audit tabulated more than 231,000 cards
    (the expected BLCA sample size was $\approx$103,000 cards); ONEAudit would have audited $\approx$1,300 cards.
    For data from a pilot hybrid RLA in Kalamazoo, MI, in 2018, ONEAudit gives a
    risk of 2\%, substantially lower than the 3.7\% measured risk for SUITE, the ``hybrid'' method the pilot used.
\end{abstract}


\vspace{0.2in}
\noindent
\textbf{An abridged version of this paper will appear in Proceedings of Voting '23, 8th Workshop on Advances in Secure Electronic Voting.}

\section{Introduction: Efficient Risk-Limiting Audits}
A procedure is a risk-limiting audit (RLA) with risk limit $\alpha$ if it guarantees that if the reported outcome is right,
the procedure will not change it; but if the reported outcome is wrong,
the chance the procedure will not correct it---the ``risk''---is at most $\alpha$.
``Outcome'' means who or what won, not the precise vote tallies.
RLA methods have been developed for many sampling designs
\cite{stark08a,stark08d,ottoboniEtal18,ottoboniEtal20,stark20,stark23b},
and to use the audit data in different ways to measure ``risk'' \cite{stark08a,stark08d,stark09b,stark10d,lindemanEtal12,lindemanStark12,stark20,stark23b},
to accommodate legal and logistical constraints and heterogeneous equipment within and across
jurisdictions.

``Card'' or ``ballot card'' means a sheet of paper; a ballot comprises one or more cards.
A ``cast-vote record'' (CVR) is the voting system's interpretation of the votes on a particular card.
A ``manual-vote record'' (MVR) is the auditors' reading of the votes on a card.
The main approaches to RLAs are \emph{ballot-polling RLAs} (BPA), which examine individual randomly selected 
cards but do not use data from the voting system other than the totals; 
\emph{batch-level comparison RLAs} (BLCA), which compare reported vote subtotals for randomly selected
batches of cards (e.g., all cards cast in a precinct) to manual tabulations of the same batches; 
 \emph{card-level comparison RLAs} (CLCA), which compare individual CVRs to the corresponding MVRs
for a random sample of cards;
and \emph{hybrid} audits, which combine two or more of the approaches above.

BLCA is closest to historical statutory audits, but requires larger sample sizes than other methods when outcomes
are correct.
BPA requires almost no data from the voting system.
It is generally more efficient than BLCA, but its sample size grows approximately quadratically as the margin shrinks.
The most efficient approach is CLCA, for which the sample size grows approximately linearly as the
margin shrinks.
But it requires the most information from the voting system: an exported CVR for every card
and a way to link exported CVRs to the corresponding physical cards, without compromising voter privacy.

This paper shows that applying CLCA to any combination of CVRs provided
by the voting system and CVRs created by the auditors to match batch subtotals or contest totals gives a valid RLA.
When the CVRs are derived entirely from contest totals, the method is algebraically equivalent to BPA.
When the CVRs are derived from batch subtotals, the method is far more efficient than BLCA
and approaches the efficiency of `pure' CLCA when the batches are sufficiently homogeneous.

Many modern voting systems can provide linked CVRs for some ballot cards (e.g., vote-by-mail)
but not others (e.g., in-precinct voters).
This has led to a variety of strategies:
\begin{itemize} 
\item give up the efficiency of CLCA and use BPA
\item
\emph{hybrid} RLAs that use different audit strategies in different strata \cite{ottoboniEtal18,stark20,stark23b,spertusStark22}
\item
BLCA using weighted random samples \cite{stark10d,lindemanEtal18,ottoboniEtal18},
with batches of size 1 for cards with linked CVRs
\item
CLCA that rescans some or all of the cards to create linked CVRs
the voting system did not originally provide \cite{rhodeIslandBrennan19,harrisonEtal22}
\item
CLCA using cryptographic nonces to link CVRs to cards without compromising voter privacy
\cite{stark23a}.
\end{itemize}
Section~\ref{sec:hetero} develops a simpler approach
that in examples is more efficient than a hybrid audit or BLCA, works even when a BCLA is impracticable,
avoids the expense of re-scanning any ballots, and does not require
new or additional equipment.
When reporting batches are sufficiently homogeneous, the sample size for the method approaches that of CLCA.

\section{Testing net overstatement does not require CVRs linked to ballot cards} \label{sec:overEasy}

\subsection{Warmup: 2-candidate plurality contest}
Consider a two-candidate plurality contest, Alice v. Bob, with Alice the reported winner.
We encode votes and reported votes as follows.
There are $N$ cards.
Let $b_i = 1$ if card $i$ has a vote for Alice, $-1$ if it has a vote for Bob, and $0$ otherwise.
Let $c_i = 1$ if card $i$ was counted by the voting system as a vote for Alice, $c_i=-1$ if it was counted as a vote for 
Bob, and $c_i=0$ otherwise.
The true margin is $\sum_{i=1}^N b_i$ and the reported margin is $\sum_{i=1}^N c_i$. 
The \emph{overstatement} of the margin on the $i$th card is $c_i-b_i \in \{-2, -1, 0, 1, 2\}$.
It is the number of votes by which the voting system exaggerated the number of votes for Alice.
Alice really won if the \emph{net overstatement of the margin}, $ E(\{c_i\}) := \sum_i (c_i - b_i)$, is 
less than the reported margin $\sum_i c_i$.
Because addition is commutative and associative,
if $\{ c_i \}$ and $\{c'_i \}$ are any two sets of CVRs for which $\sum_i c_i = \sum_i c'_i$, then $E(\{c_i\}) = E(\{c'_i\})$:
they are \emph{overstatement net equivalent} (ONE).
\begin{equation}
E(\{c_i\}) := \sum_i (c_i - b_i) =  \sum_i c_i - \sum_i b_i = \sum_i c'_i - \sum_i b_i = \sum_i (c'_i - b_i) =  E(\{c'_i\}).
\end{equation}
Hence, if we have an RLA procedure to test whether $E(\{c_i\}) < \sum_i c_i$ using the ``real'' CVRs produced by the voting 
system, the same 
procedure can test whether the outcome is correct if it is applied to other CVRs $\{c'_i\}$ provided $\sum_i c_i = \sum_i c'_i$, 
\emph{even if the CVRs $\{c'_i\}$ did not come from the voting system}.
(Audit sample sizes might be quite different.)

Thus, we can conduct a CLCA using \emph{any} set $\{c'_i\}$ of CVRs that reproduces the contest-level results:
the net overstatement of every such set of CVRs is the same.
If the system reports batch-level results, we can require that the CVRs reproduce the batch-level results as well.,
which might reduce audit sample sizes, especially when the batches have different political preferences.
If the voting system reports CVRs for some individual ballot cards, we can conduct a CLCA that uses those CVRs,
augmented by ONE CVRs for the remaining ballot cards (derived from batch subtotals or from contest totals, by subtraction). 
Better agreement between the MVRs and VCRs will generally allow the audit to stop after inspecting fewer cards. 

\subsection{Numerical example}
Consider a contest in which 20,000 cards were cast in all, of which 10,000 were cast by mail and 
have linked CVRs, with 
5,000 votes for Alice, 4,000 for Bob, and 1,000 undervotes.
The other 10,000 cards were cast in 10~precincts, 1,000 cards in each.
Net across those 10~precincts, Alice and Bob each got 5,000 votes.
In 5~precincts, Alice showed more votes than Bob; in the other 5, Bob showed more than Alice.
The reported results are thus 10,000 votes for Alice, 9,000 for Bob, and 1,000 undervotes.
The margin is 1,000 votes; the \emph{diluted margin} (margin in votes, divided by cards cast) is $1000/20000 = 5\%$.
Consider two sets of precinct subtotals:
\begin{itemize}
   \item 5 precincts show 900 votes for Alice and 100 for Bob; the other 5 show 900 votes for Bob and
 100 for Alice.
   \item 5 precincts show 990 votes for Alice and 10 for Bob; the other 5 show 990 votes for Bob and 10 for Alice.
\end{itemize}
Construct ONE CVRs for the 10,000 cards cast in the 10 precincts as follows:
if the precinct reports $a$ votes for Alice and $1000-a$ for Bob, the net vote for Alice is $a - (1000-a) = 2a-1000$.
The ``average'' CVR for the precinct has $(2a-1000)/1000 = 2a/1000 - 1$ votes for Alice; that is the ONEAudit CVR for every
card in the precinct.
For instance,  a precinct that reported 900 votes for Alice and 100 for Bob has a net margin of $900\times1 + 100 \times -1 = 800$
for Alice, so that precinct contributes 1,000 ONE CVRs, each with $c_i = (0.9)\times 1 + (0.1)\times(-1) = 0.8$ votes for Alice. 
\begin{SCtable}
\centering
\begin{tabular}{lr}
\hline
batch total  & ONE CVR \\
\hline
990 Alice, 10 Bob & 0.98 Alice \\
900 Alice, 100 Bob & 0.8  Alice \\
100 Alice, 900 Bob & -0.8 Alice (0.8 Bob) \\
10 Alice, 990 Bob & -0.98 Alice (0.98 Bob)
\end{tabular}
\caption{\protect \label{tab:one-cvr}
Overstatement-net equivalent CVRs that match batch subtotals.
If a precinct of 1000 voters reported 990 votes for Alice and 10 for Bob, the net overstatement is the
same as if there had been 1,000 CVRs, one for each card, each showing 0.98 votes for Alice.}
\end{SCtable}

To audit, draw ballot cards uniformly at random, without replacement. 
To find the overstatement for each audited card, subtract
the MVR for the card (-1, 0, or 1) from the CVR (-1, 0, or  1) if the system provided one, or 
from the ONE CVR for its precinct (a number in $[-1, 1]$) if the system did not provide a CVR.
Apply a ``risk-measuring function'' (see, e.g., \cite{stark20,stark23b}) to the overstatements
to measure the risk that the outcome is wrong based on the data collected so far; the audit can stop without a full 
hand count if and when the measured risk is less than or equal to the risk limit.

The random selection can be conducted in many ways, for instance, conceptually numbering the cards from
1 to 20,000, where cards 1--10,000 are the ballots with CVRs, ordered in some canonical way;
cards 10,001--11,000 are the cards cast in precinct~1, starting with the top card in the stack;
cards 11,001--12,000 are the cards cast in precinct~2, starting with the top card in the stack; etc.
Auditors draw random numbers between 1 and 20,000, and retrieve the corresponding card.
Alternatively, if the resulting number is between 1 and 10,000, retrieve the corresponding card;
but if the number is larger, draw a ballot at random from the precinct that numbered card belongs to, for instance, 
using the $k$-cut method \cite{sridharRivest20}. 
That approach avoids counting into large stacks of ballots.

Table~\ref{tab:hetero} summarizes expected audit sample sizes.
If there had been a CVR for every card and the results were exactly correct, the sample size for
a standard CLCA with risk limit 5\% would be about 125 cards.
A BPA at risk limit 5\% would examine about 2,300 cards.
A BLCA (treating individual cards as batches for those with CVRs) using sampling with probability
proportional to an error bound and the Kaplan-Markov test 
\cite{stark09b} would examine
about 7250 cards on average in the 900/100 scenario and 5300 in the 990/10 scenario.
For ONEAudit,
the expected sample size is about 800 cards in the 900/100 scenario and 170 in the 990/10 scenario.
As preferences within precincts become more homogeneous, ONEAudit approaches the efficiency of CLCA.
\begin{table}[ht]
\centering
\begin{tabular}{llrrr}
\hline
      &               & expected &  & \\
scenario & method &  cards & vs BPA &  vs CLCA  \\
\hline
both & BPA &  2,300 & 1.00  & 18.40 \\
        & CLCA & 125  & 0.05 & 1.00  \\
\hline
900/100 & BLCA & 7,250 & 3.15 &  58.00 \\
& ONE CLCA  & 800 & 0.35 & 6.40 \\
\hline
990/10 & BLCA & 5,300 & 2.30 & 42.40 \\
            & ONE CLCA& 170 & 0.07 & 1.36
\end{tabular}
\caption{\protect \label{tab:hetero} Expected RLA workloads for a two-candidate
plurality contest with a `diluted margin' of 5\% at risk limit 5\%.
In all, 20,000 cards were cast, of which 10,000 have linked CVRs; the other 10,000 cards are 
divided into 10~precincts with 1,000 cards each.
Among cards with CVRs, 5,000 show votes for the winner, 4,000 show votes for the loser, and 1,000 have invalid votes.
In the 900/100 scenario, 5~precincts have 900 votes for the winner
and 100 for the loser; the other 5 have 900 for the loser and 100 for the winner.
In the 990/10 scenario, 5~precincts have 990 votes for the winner
and 10 for the loser; the other 5 have 990 for the loser and 10 for the winner.
BPA: ballot-polling audit.
CLCA: card-level comparison audit (which would require re-scanning the 10,000 cards cast in precincts).
BLCA: batch-level comparison audit.
ONE CLCA: card-level comparison audit using the original 10,000 CVRs, augmented by
10,000 ONE CVRs derived from precinct subtotals.
Column~4: sample size divided by BPA sample size.
Column~5: sample size divided by CLCA sample size.
Expected workload for BPA and CLCA is the same in both scenarios. 
Sample sizes for CLCA use the ``super-simple'' method \cite{stark10d},
computed using \url{https://www.stat.berkeley.edu/~stark/Vote/auditTools.htm} (last visited 
19~March 2023).
Sample sizes for BPA use ALPHA \cite{stark23b}.
}
\end{table}

\subsection{The general case}

We use the SHANGRLA audit framework \cite{stark20} because it can be works with
every social choice function for which an RLA method is known; however, ONE CVRs can be used with 
every extant RLA method for comparison audits.
SHANRGLA reduces auditing election outcomes to multiple instances of a single problem: testing 
whether the mean of a finite list of bounded numbers is less than or equal to $1/2$.
Each list results from applying a function $A$ called an \emph{assorter} to the votes on the ballots; each assorter
maps votes to the interval $[0, u]$, where the upper bound $u$ depends on the particular assorter.
Different social choice functions involve different assorters and, in general, different numbers of assorters.
An \emph{assertion} is the claim that the average of the values the assorter takes on the true votes 
is greater than $1/2$.
The contest outcome is correct if the all the SHANGRLA assertions for the contest are true.

A \emph{reported assorter margin} is the amount by which that assorter applied to the
reported votes exceeds $1/2$.
For scoring rules (plurality, supermajority, multi-winner plurality,
approval voting, Borda count, etc.), reported assorter margins can be computed from contest-level tallies, batch tallies, or CVRs.
For auditing IRV using RAIRE \cite{blomEtal18}, CVRs are generally required to construct an
appropriate set of assorters and to find their margins---but the CVRs do not have to
be linked to individual ballot cards.

Let $b_i$ denote the true votes on the $i$th ballot card; there are $N$ cards in all.
Let $c_i$ denote the voting system's interpretation of the $i$th card. 
Suppose we have a CVR $c_i$ for every ballot card whose index $i$ is in $\cC$.
The cardinality of $\cC$ is $ |\cC|$.
Ballot cards not in $\cC$ are partitioned into $G \ge 1$ disjoint groups $\{\cG_g\}_{g=1}^G$ for which
reported assorter subtotals are available.
For instance $\cG_g$ might comprise all ballots for which no CVR is available or all
ballots cast in a particular precinct.
Unadorned overbars denote the average of a quantity across all $N$ ballot cards; overbars subscripted by a set (e.g., $\cG_g$) 
denote the average of a quantity across cards in that set, for instance:
$$
 \bar{A}^c := \frac{1}{N} \sum_{i=1}^N A(c_i), \;\;\;
 \bar{A}_\cC^c := \frac{1}{|\cC|} \sum_{i \in \cC } A(c_i), \;\;\;
 \bar{A}_{\cG_g}^c := \frac{1}{|\cG_g|} \sum_{i \in \cG_g } A(c_i)  
$$
$$
 \bar{A}^b := \frac{1}{N} \sum_{i=1}^N A(b_i),  \;\;\;
 \bar{A}_\cC^b := \frac{1}{|\cC|} \sum_{i \in \cC } A(b_i), \;\;\;
 \bar{A}_{\cG_g}^b := \frac{1}{|\cG_g|} \sum_{i \in \cG_g } A(b_i). 
$$
The \emph{assertion} is the claim $\bar{A}^b > 1/2$.
The reported assorter mean $\bar{A}^c  > 1/2$: otherwise, according to the voting system's data,
the reported outcome is wrong. 
Now $\bar{A}^b > 1/2$ iff
\beq \label{eq:difference_condition}
\bar{A}^c - \bar{A}^b < \bar{A}^c-1/2.
\eeq
The right hand side is known before the audit starts; it is half the ``reported assorter margin'' $v := 2\bar{A}^c-1$ \cite{stark20}.
We assume we have a \emph{reported assorter total} 
$\sum_{i \in \cG_g} A(c_i)$ from the voting system for the cards in the group $\cG_g$ 
(e.g., reported precinct subtotals) and define the \emph{reported assorter mean} for $\cG_g$:
\beq
\hat{A}_{\cG_g}^c :=  \frac{1}{|\cG_g|} \sum_{i \in \cG_g} A(c_i).
\eeq
We have 
\beq
    \bar{A}^c = \frac{\sum_{g=1}^G |\cG_g| \hat{A}_{\cG_g}^c  + \sum_{i \in \cC} A(c_i)}{N} = 
      \frac{\sum_{g=1}^G \sum_{i \in \cG_g} \hat{A}_{\cG_g}^c  + \sum_{i \in \cC} A(c_i)}{N}. \label{eq:basic-identity}
\eeq
Thus if we \emph{declare} $A(c_i) := \hat{A}_{\cG_g}^c$ for  $i \in \cG_g$, the reported assorter mean for the cards in group
$\cG_g$, the mean of the assorter applied to the CVRs---including these faux CVRs---will equal its reported value:
using a ``mean CVR'' for the batch is overstatement-net-equivalent to any CVRs that give the same
assorter batch subtotals. 
Condition \ref{eq:difference_condition} then can be written
\begin{equation}
\frac{1}{N} \sum_i (A(c_i)-A(b_i))  < v/2.
\eeq
Following SHANGRLA \cite[section 3.2]{stark20}, define
\beq
B(b_i)  :=  \frac{u + A(b_i)-A(c_i)}{2u-v} \in [0, 2u/(2u-v)].
\eeq
Then $\bar{A}^b > 1/2 \;\; \iff \;\; \bar{B}^b > 1/2$,
which can be shown as follows, using the fact that $v := 2\bar{A}^c-1 \le 2u-1 < 2u$:
\begin{eqnarray}
\bar{B}^b &:=& \frac{1}{N}\sum_i \frac{u + A(b_i)-A(c_i)}{2u-v} \nonumber \\
&=& \frac{u + \bar{A}^b - \bar{A}^c}{2u-v} \nonumber \\
&=& \frac{u + \bar{A}^b - \bar{A}^c}{2u-2\bar{A}^c + 1}.
\end{eqnarray}
Thus if $\bar{B}^b > 1/2$,
\begin{eqnarray}
\frac{u+ \bar{A}^b - \bar{A}^c}{ 2u - 2\bar{A}^c +1} & > &1/2 \nonumber \\
u + \bar{A}^b - \bar{A}^c & > & u - \bar{A}^c + 1/2   \nonumber \\
 \bar{A}^b & > & 1/2.
\end{eqnarray}
If the reported tallies are correct, i.e., if $\bar{A}^c = \bar{A}^b = (v+1)/2$, then 
\beq
   \bar{B}^b = \frac{u}{2u-v}.
\eeq

\section{Auditing using batch subtotals}
The oldest approach to RLAs is batch-level comparison, which involves exporting batch subtotals from the voting system
(e.g., for precincts or tabulators), verifying that those batch subtotals yield the reported contest results,
drawing some number of batches at random (with equal probability or with probability proportional to an error bound), manually tabulating
all the votes in each selected batch, comparing the
manual tabulation to the reported batch subtotals, assessing whether
the data give sufficiently strong evidence that the reported results are right, and expanding the sample if not \cite{stark08a,stark08d,stark08c,stark09b,higginsEtal11}.

BLCAs have two logistical hurdles: 
(i)~They require manually tabulating the votes on \emph{every} ballot card in the batches selected for audit.
(ii)~When the batches of cards for which the voting system reports
subtotals do not correspond to identifiable physical batches (common for vote-by-mail and vote centers), the audit has to find and retrieve every card in 
the audited reporting batches. Those cards may be spread across any number of physical batches,
which can also make recounts prohibitively expensive \cite{appel23}.

Both can be avoided using CLCA with ONE CVRs. 
The following algorithm gives a valid RLA, but selects and compares the manual interpretation of 
\emph{individual} cards 
to the implied ``average'' CVR of the reporting batch each card belongs to.
We assume that the canvass and a \emph{compliance audit} \cite{appelStark20}
have determined that the ballot manifest and physical cards are complete and trustworthy.
\begin{mdframed}
\textbf{Algorithm for a CLCA using ONE CVRs from batch subtotals.}
\begin{enumerate}
    \item Pick the risk limit for each contest under audit.
    \item Export batch subtotals from the voting system.
    \item Verify that every physical card is accounted for,\footnote{%
    For techniques to deal with missing cards, see \cite{banuelosStark12,stark20}.}
     that the physical accounting is consistent with the reported votes, and that
              the reported batch subtotals produce the reported winners.
    \item Construct SHANGRLA assorters for every contest under audit; select a risk-measuring function for each assertion
    (e.g., one in \cite{stark23b});
    set the measured risk for each assertion to 1.
    \item Calculate the reported mean assorter values for each reporting batch; these are the ONE CVRs.
    \item While any measured risk is greater than its risk limit and not every card has been audited:
    \begin{itemize}
        \item Select a card at random, e.g., by selecting a batch at random with probability proportional to the size of the batch, then selecting a card uniformly at random from the batch using the $k$-cut method \cite{sridharRivest20}, or by selecting at random from the entire collection of cards.
        \item Calculate the overstatement for the selected card using the ONE CVR for the reporting batch the card belongs to.
        \item Update the measured risk of any assertion whose measured risk is still greater than its risk limit.
        \item If the measured risk for every assertion is less than or equal to its risk limit, stop and confirm the reported outcomes.
    \end{itemize}
    \item Report the correct contest outcomes: every card has been manually interpreted.
\end{enumerate}
\end{mdframed}
This algorithm be made more efficient statistically and logistically in a variety of ways, for instance, by making an affine translation of the 
data so that the minimum possible value is 0 (by subtracting the minimum of the possible overstatement assorters across batches 
and re-scaling so that the null mean is still 1/2) and by starting with a sample size that is expected to be large enough to
confirm the contest outcome if the reported results are correct.

\subsection{Numerical case studies}
Table~\ref{tab:batch} compares expected sample sizes for BLCA to CLCA using ONE CVRs derived from the same batch
subtotals, and to BPA, for two contests: the 2022 midterm Georgia Secretary of State's contest, which had a 
diluted margin (margin in votes divided by cards cast) of about 9.2\%\footnote{%
  \url{https://sos.ga.gov/news/georgias-2022-statewide-risk-limiting-audit-confirms-results}, last visited 26~February 2023.
}
and the 2020 presidential election in California, which had a diluted margin of about 28.7\%.
The Georgia contest was audited using batch-level comparisons.
The Georgia SoS claims that audit was a BLCA with a risk limit of 5\%, but
in fact the audit was not an RLA, for a variety of reasons.\footnote{%
  \url{https://www.stat/berkeley.edu/~stark/Preprints/cgg-rept-10.pdf}, last visited 15~December 2022.
}
BPA and CLCA using ONE CVRs are generally much more efficient than BLCA when batches are large.
CLCA with ONE CVRs is more efficient than
BPA when batches are more homogenous than the contest votes as a whole, i.e., when precincts are polarized in different
directions.
\begin{table}[ht]
\begin{tabular}{l|r|r|r|r|r|r|r}
Contest & total & total       & actual          &  K-M      & ONE      & Wald   & BLCA/\\
              & turnout & batches & sample size &  BLCA   & CLCA     &  BP             & CLCA\\
\hline
2020 CA U.S.\ Pres & 17,785,667 & 21,346 & $\approx$178,000 & 26,700 & 70 & 72 & 381\\
2022 GA SoS          & 3,909,983 & 12,968 & $>$231,000 & 103,300 & 1,380  & 700 & 75
\end{tabular}
\vspace{0.05in}
\caption{\protect \label{tab:batch} Actual and estimated expected sample sizes for various RLA
methods for the 2020 U.S. presidential election in California and the 2022 Georgia Secretary of State contest,
at risk limit 5\%.
Columns~2, 3: turnout per state records.
CA data from \url{https://statewidedatabase.org/d10/g20.html} (last visited 2~March 2020); the
CA Statement of Vote gives slightly smaller turnout 17,785,151 (\url{https://elections.cdn.sos.ca.gov/sov/2020-general/sov/complete-sov.pdf}, last visited 2~March 2023).
GA data from (\url{https://sos.ga.gov/news/georgias-2022-statewide-risk-limiting-audit-confirms-results}, last visited 26~February 2023).
Column~4: approximate number of cards examined in the actual batch-level audits (which were not RLAs).
Column~5: expected sample size for BLCA using the
Kaplan-Markov risk function.
Column~6: expected sample size for CLCA using ONE CVRs based on batch subtotals, 
for the ALPHA risk-measuring function with the truncated 
shrinkage estimator with parameters $c=1/2$, $d=10$, estimated from 100 Monte Carlo replications.
Column~7: expected sample size for BPA using Wald's SPRT.
Column~8: column~5 divided by column~6.
A different risk function should reduce the GA ONE CLCA sample size to at most the BPA sample size:
see section~\ref{sec:theory}.
Estimates assume that reported batch subtotals are correct. 
}
\end{table}

\section{Auditing heterogenous voting systems} \label{sec:hetero}
When the voting system can report linked CVRs for some but not all cards, we can
augment the voting system's linked CVRs with ONE CVRs for the remaining cards, then use CLCA.
The ONE CVRs can be derived from overall contest results or from reported subtotals, e.g., precinct subtotals.
Finer-grained subtotals generally give smaller audit sample sizes (when the reported outcome is correct) 
if the smaller groups are more homogeneous than the overall election.

SUITE \cite{ottoboniEtal18}, a hybrid RLA designed for this situation, was first fielded in a pilot audit of the gubernatorial 
primary in Kalamazoo, MI, in 2018.
The stratum with linked CVRs comprised 5,294 ballots with 5,218 reported votes in the contest; the ``no-CVR'' stratum
comprised 22,372 ballots with 22,082 reported votes.
The sample included 32 cards were drawn from the no-CVR stratum and 8~from the CVR stratum.\footnote{%
See \url{https://github.com/kellieotto/mirla18/blob/master/code/kalamazoo_SUITE.ipynb}.
}
Table~\ref{tab:kalamazooPoll} summarizes the contest and audit results; auditors found no errors
in the 8~CVRs, each of which yields the overstatement assorter value $u/(2u-v)$.
The new method compares the 32~cards without CVRs to ONE CVRs derived from all the votes without CVRs
(not from subtotals for smaller batches) by subtracting the votes on the linked CVRs from the reported contest totals.
Ignoring the fact that the sample was stratified and the difference in sampling fractions in the two strata, in 100,000
random permutations of the data, the ALPHA martingale test using a fixed alternative
 $0.99 (2u)/(2u-v)$ had a mean $P$-value of $0.0201$
 (90th percentile $0.0321$), about 54\%
 of the SUITE $P$-value of $0.0374$ \cite{ottoboniEtal18}.
 The ONEAudit $P$-value is larger than the $P$-value of the best product supermartingale test 
 in \cite{spertusStark22}, but comparable to or smaller than the $P$-values for the other supermartingale tests.
 See table~\ref{tab:compare}.
 If precinct subtotals were available to construct the ONE CVRs, the measured risk might have been
 lower, depending on precinct heterogeneity.

\begin{SCtable}[][ht]
\centering{
\begin{tabular}{lrr|r}
\hline
Candidate & CVR            & no-CVR  &   polling \\
                 & & & sample  \\
\hline
Butkovich &  6 & 66 & 0 \\
Gelineau &  56 & 462 & 1 \\
Kurland &    23 & 284 & 0 \\
Schleiger &  19 & 116 & 0 \\
Schuette &  1,349 &  4,220 & 8 \\
Whitmer &  3,765 & 16,934 & 23 \\
Non-vote & 76 & 290 & 0 \\
\hline
Total & 5,294 & 22,372 & 32
\end{tabular}
}
\caption{\protect \label{tab:kalamazooPoll} Reported votes in the stratum with CVRs and the stratum without CVRs,
and the audited votes in the random sample of 32~cards from 
the stratum without CVRs in the 2018 RLA pilot in Kalamazoo, MI.}
\end{SCtable}

\begin{table}[ht]
\centering{
\begin{tabular}{lrrr}
\hline
Method & $P$-value & SD & 90th \\
             &         &       & percentile \\
\hline
SUITE & 0.037 & n/a & n/a \\
ALPHA $P^*_F$ & 0.018 & 0.002 & 0.019 \\
ALPHA $P^*_M$ & 0.003 & 0.000 & 0.003 \\
Empirical Bernstein $P^*_F$ & 0.348 & 0.042 & 0.390 \\
Empirical Bernstein $P^*_M$ & 0.420 & 0.134 & 0.561 \\
ALPHA ONEAudit & 0.020 & 0.010 & 0.032 
\end{tabular}
\caption{\protect \label{tab:compare} $P$-values for the 2018 RLA pilot in Kalamazoo, MI, for
different risk-measuring functions.
SUITE is a hybrid stratified approach \cite{ottoboniEtal18}.
Rows 2--5 are from \cite[Table 3]{spertusStark22}: the ALPHA and Empirical Bernstein stratumwise supermartingales
combined using either Fisher's combining function
($P^*_F$) or multiplication ($P^*_M$).
The 6th row is for ONEAudit, using the ALPHA supermartingale with fixed alternative $\eta = 0.99$,
a non-adaptive choice.
Technically, ONEAudit should not be applied to this sample because the sample was stratified, while the risk calculation
assumes the sample was a simple random sample of ballot cards:
this is just a numerical illustration.}
}
\end{table}

\section{Sample sizes for contest-level ONE CLCA vs. BPA}
\subsection{Theory} \label{sec:theory}
Moving from tests about raw assorter values to tests about overstatements relative to ONE CVRs derived
from overall contest totals is just an affine transformation: no information is gained or lost.
Thus, if we audited using an affine equivariant statistical test, the sample size should be the same whether
the data are the original assorter values (i.e., BPA) or overstatements from ONE CVRs.

However, the statistical tests used in RLAs are not affine equivariant because they rely on \emph{a priori} bounds
on the assorter values.
The original assorter values will generally be closer to the endpoints of $[0, u]$ than the transformed values are
to the endpoints of $[0, 2u/(2u-v)]$.
To see why, suppose that there are no reported CVRs ($\cC = \emptyset$) and that only contest
totals are reported from the system---so every cast ballot card is in $\cG_1$.
For a BPA, the population values from which the sample is drawn are the original assorter values $\{A(b_i)\}$,
which for many social choice functions can take only the values $0$, $1/2$, and $u$.
For instance, consider a two-candidate plurality contest, Alice v. Bob, where Alice is the reported winner.
This can be audited using a single assorter that
assigns the value $0$ to a card with a vote for Bob, the value $u=1$ to a card with a vote for Alice, and the value $1/2$ to other cards.
In contrast, for a comparison audit, the possible population values $\{B(b_i)\}$ are
$$ \left \{\frac{1 - (v+1)/2}{2-v}, 1/2, \frac{2 - (v+1)/2}{2-v} \right \}.
$$
Unless $v=1$---i.e., unless every card was reported to have a vote for Alice---the minimum value of the overstatement 
assorter is greater than $0$ and the maximum is less than $u$.
Figure~\ref{fig:extrema_function_v} plots the minimum and maximum value of the overstatement assorter
as a function of $v$ for $u=1$.
\begin{SCfigure}
\centering
\includegraphics[scale=0.35]{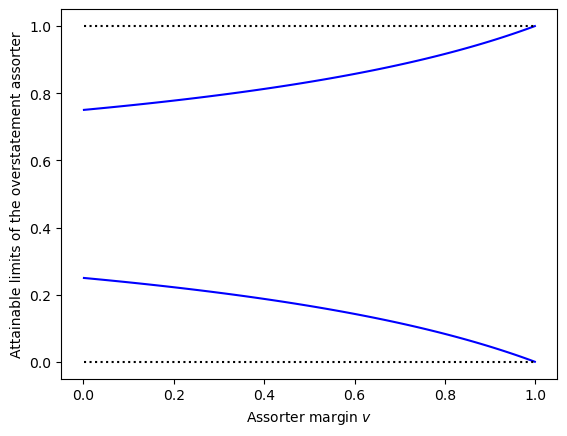}
\caption{ \protect \label{fig:extrema_function_v} Upper and lower bounds on the overstatement assorter as a function of the diluted margin $v$,
for $u=1$.}
\end{SCfigure}

A test that uses the prior information $x_j \in [0, u]$ may not be as efficient for populations for which $x_j \in [a, b]$ with
$a>0$ and $b<u$ as it is for populations where the values $0$ and $u$ actually occur.
An affine transformation of the overstatement assorter values can move them back to the endpoints
of the support constraint by subtracting the minimum possible value
then re-scaling so that the null mean is $1/2$ once again, which reproduces the original assorter, $A$:
\begin{eqnarray}
   C(b_i) &:=& \frac{1/2}{1/2-\frac{u - (v+1)/2}{2u-v}}  \cdot \left ( B(b_i) - \frac{u - (v+1)/2}{2u-v} \right )  \nonumber \\
   &=& \frac{2u-v}{2u-v -(2u - (v+1))} \cdot \left (\frac{u + A(b_i) - (v+1)/2}{2u-v}  -  \frac{u - (v+1)/2}{2u-v} \right )  \nonumber \\
   &=& (2u - v) \cdot \frac{A(b_i)}{2u-v}  =  A(b_i).
\end{eqnarray}

\noshow{One way to increase power is to use the \emph{taint transformation}, which expresses overstatements
as a fraction of the maximum possible overstatement on that card (or group of cards, for batch-level audits), then samples
each card (or group of cards) with probability proportional to its maximum possible overstatement (probability proportional to
error bound sampling, PPEB).
This is the most efficient known approach to batch-level comparison audits \cite{stark09b}.
The upper bound on every taint is 1. 
For SHANGRLA assorters, the maximum overstatement is the reported assorter value for the card (or the reported assorter sum, for groups of cards).
}

\subsection{Numerical comparison}
While CLCA with ONE CVRs is algebraically equivalent to BPA, the performance of a given statistical
test will be different for the two formulations. 
We now compare expected audit sample sizes for some common risk-measuring functions applied to
CLCA with ONE CVRs derived from contest-level results and applied to the original assorter data.
This is assessing the particular \emph{statistical tests}, not 
any intrinsic difference between BPA and ONE CLCA.

Tables~\ref{tab:performance-detail-1}--\ref{tab:performance-detail-5} in the appendix
report expected sample sizes when the
reported winner received a share $\theta \in \{0.505, 0.51, 0.52, 0.55, 0.6\}$ of the reported votes,
for percentages of cards that do not contain a valid vote for either candidate ranging from 10\% to 75\%, 
and various values of tuning parameters in the risk-measuring functions.
Table~\ref{tab:summary} gives the geometric mean of
the ratios of the mean sample size for each condition to the smallest mean sample size for that condition (across risk-measuring functions).
Transforming the assorter into an overstatement assorter using the ONEAudit transformation, then testing whether the mean
of the resulting population is $\le 1/2$ using the ALPHA test martingale with the 
truncated shrinkage estimator of \cite{stark23b} with
$d=10$ and $\eta$ between $0.505$ and $0.55$ performed comparably to---but slightly worse than---using ALPHA
on the raw assorter values for the same $d$ and $\eta$, and within 4.8\% of the overall performance of the best-performing
method.
\begin{SCtable}
\tiny
\begin{tabular}{llr|lr}
Method & Parameters & Score & Method & Score \\
\hline ALPHA & $\eta=$0.505 $d=$10 & 1.51  &  ONEAudit & 1.52 \\
 & $\eta=$0.505 $d=$100 & 1.54  &  & 1.55 \\
 & $\eta=$0.505 $d=$1000 & 1.79  &  & 1.83 \\ 
 & $\eta=$0.505 $d=\infty$ & 3.02  & & 3.83 \\
\cline{2-5}
 & $\eta=$0.51 $d=$10 & 1.51  &  & 1.52 \\ 
 & $\eta=$0.51 $d=$100 & 1.53  &  & 1.55 \\ 
 & $\eta=$0.51 $d=$1000 & 1.72  &  & 1.80 \\ 
 & $\eta=$0.51 $d=\infty$ & 2.29  & & 3.05 \\
\cline{2-5}
 & $\eta=$0.52 $d=$10 & 1.51  &  & 1.53 \\ 
 & $\eta=$0.52 $d=$100 & 1.51  &  & 1.55 \\ 
 & $\eta=$0.52 $d=$1000 & 1.61  &  & 1.75 \\ 
 & $\eta=$0.52 $d=\infty$ & 1.84  & & 2.32 \\
\cline{2-5}
 & $\eta=$0.55 $d=$10 & 1.51  &  & 1.55 \\ 
 & $\eta=$0.55 $d=$100 & 1.47  &  & 1.56 \\ 
 & $\eta=$0.55 $d=$1000 & \bf{1.44}  &  & 1.62 \\  
 & $\eta=$0.55 $d=\infty$ & 1.88  &  & 1.81 \\
\cline{2-5}
 & $\eta=$0.6 $d=$10 & 1.50  & & 1.59 \\ 
 & $\eta=$0.6 $d=$100 & 1.45  & & 1.57 \\  
 & $\eta=$0.6 $d=$1000 & 1.51  & & 1.52 \\  
 & $\eta=$0.6 $d=\infty$ & 2.42  & & 1.84 \\
\cline{2-5}
\hline SqKelly & & 1.98 \\ 
 \hline a priori Kelly 
 & $\eta=$0.505 & 2.77 \\
 & $\eta=$0.51 & 1.88 \\
 & $\eta=$0.52 & 1.60 \\
 & $\eta=$0.55 & 2.14 \\
 & $\eta=$0.6 & 3.34 \\
\end{tabular}
\caption{\protect \label{tab:summary} Geometric mean of the ratios of sample sizes to the smallest sample size
for each condition described above.
The smallest ratio is in bold font.
In the simulations, for the hypothesis tests considered, the ONEAudit transformation entails a negligible loss in efficiency
compared to ALPHA applied to the ``raw'' assorter.}
\end{SCtable}

\section{Conclusions}
Ballot-polling risk-limiting audits (BPAs) are algebraically equivalent to card-level comparison risk-limiting audits (CLCAs) using 
faux cast-vote records (CVRs) that match the overall reported results. 
Any set of CVRs that reproduces the reported contest tallies (more generally, reproduces reported assorter totals) has the same net 
overstatement of the margin: 
they are ``overstatement-net-equivalent'' (ONE) to the
voting system's tabulation and can be used as if the voting system had exported them.

ONE CVRs let audits use batch-level data far more efficiently than traditional batch-level 
comparison RLAs (BLCAs) do: create ONE CVRs for each batch, then apply CLCA as if the voting system had provided those CVRs. 
For real and simulated data, this saves a large mount of work
compared to manually tabulating the votes on \emph{every}
card in the batches selected for audit, as BLCAs require.
If batches are sufficiently homogeneous, the workload approaches that of ``pure'' CLCA using linked CVRs from the voting system.
BLCAs also require locating and retrieving every card in each batch that is selected for audit.
That is straightforward when reporting batches are identifiable physical batches, but not when physical batches contain a
mixture of ballot cards from different reporting batches, which is common in jurisdictions that use vote centers or
do not sort vote-by-mail ballots before scanning them.
In the California presidential election in 2020 and the Georgia election for Secretary of State in 2022, this approach would 
reduce the workload by a factor of 75 to 380, respectively, compared to the most efficient known method for BLCA.

ONE CVRs also can obviate the need to stratify, to rescan cards, or to use ``hybrid'' audits when the 
voting system cannot export a linked CVR for every card: create ONE CVRs for the
cards that lack them, then apply CLCA as if the CVRs had been provided by the voting system.
The same CLCA software can be used to audit voting systems that can export a linked CVR for every
card and also those that cannot. 
For data from a 2018 pilot audit in Kalamazoo, MI, ONE CLCA gives a measured risk much smaller
than that of SUITE (2\% versus 3.7\%), the hybrid method used in the pilot. 
Stratification and hybrid audits increase the complexity and opacity of audits,
and rescanning substantially increases time and cost.
Hence, ONEAudit may be cheaper, faster, simpler, and more transparent than previous methods.
Software used to generate the tables and figures is available at \url{https://www.github.com/pbstark/ONEAudit}.

\paragraph*{Acknowledgments.}
This work was supported by \blinded{NSF Grant SaTC--2228884}. I am grateful to Jake Spertus and Andrew Appel for helpful conversations.

\bibliography{../../Bib/pbsBib.bib}

\appendix

\section{Detailed simulation results for BPA versus CLCA using ONE CVRs based on contest totals}

\begin{table}[ht]
\tiny
\begin{tabular}{llr|rrrr|rrrr|rrrr} 
& & & \multicolumn{4}{|c|}{$N=$10,000, \%blank} &  \multicolumn{4}{|c|}{$N=$100,000 \%blank} & \multicolumn{4}{|c}{$N=$500,000 \%blank} \\ 
$\theta$ & Method & params & 10 & 25 & 50 & 75  & 10 & 25 & 50 & 75  & 10 & 25 & 50 & 75  \\
\hline 0.505 & sqKelly & & 8,852  & 9,050  & 9,024  & 9,247  & 89,777  & 90,146  & 89,742  & 89,914  & 456,305  & 449,438  & 442,418  & 449,083  \\
\cline{2-15} & apKelly & $\eta=$0.505 & 9,590  & 9,804  & 9,928  & 9,977  & 38,995  & 43,712  & 54,354  & 75,746  & \bf{56,043}  & \bf{66,229}  & 91,978  & 153,452  \\
& ALPHA & $\eta=$0.505 $d=$10 & 8,237  & 8,635  & 9,013  & 9,771  & 46,485  & 49,763  & 59,800  & 82,477  & 84,352  & 95,843  & 137,445  & 257,907  \\
&  &  $d=$100 & 8,136  & 8,610  & 9,033  & 9,777  & 43,951  & 48,756  & 59,587  & 82,534  & 78,259  & 91,725  & 134,181  & 256,568  \\
&  &  $d=$1000 & 8,108  & 8,684  & 9,078  & 9,785  & 42,094  & 46,855  & 59,413  & 82,679  & 72,004  & 87,454  & 131,923  & 256,439  \\
 &  &     $d=\infty$ & 8,180  & 8,712  & 9,194  & 9,802  & 37,106  & 43,424  & 58,777  & 83,714  & 56,227  & 71,123  & 118,816  & 257,356  \\
& ONEAudit & $\eta=$0.505 $d=$10 & 8,232  & 8,637  & 9,015  & 9,774  & 46,408  & 49,802  & 59,895  & 82,625  & 84,159  & 95,708  & 137,878  & 258,783  \\
&  & $d=$100 & 8,146  & 8,628  & 9,042  & 9,778  & 43,869  & 48,777  & 59,697  & 82,656  & 77,946  & 91,620  & 134,828  & 257,724  \\
&  & $d=$1000 & 8,165  & 8,720  & 9,096  & 9,788  & 42,256  & 47,124  & 59,582  & 82,843  & 72,313  & 87,728  & 132,902  & 257,921  \\
 &  & $d=\infty$ & 8,405  & 8,868  & 9,279  & 9,815  & 43,941  & 50,644  & 65,198  & 85,899  & 76,545  & 97,537  & 162,012  & 299,847  \\
\cline{2-15} & apKelly & $\eta=$0.51 & 8,550  & 9,067  & 9,432  & 9,936  & 37,498  & 40,902  & 48,425  & 63,746  & 80,498  & 88,848  & 110,767  & 153,644  \\
& ALPHA & $\eta=$0.51 $d=$10 & 8,244  & 8,642  & 9,004  & 9,771  & 46,624  & 49,708  & 59,802  & 82,467  & 84,466  & 95,854  & 137,524  & 257,857  \\
&  &  $d=$100 & 8,145  & 8,599  & 9,024  & 9,771  & 44,062  & 48,808  & 59,517  & 82,466  & 78,549  & 91,878  & 134,385  & 256,440  \\
&  &  $d=$1000 & 8,027  & 8,613  & 9,044  & 9,782  & 42,099  & 46,647  & 59,140  & 82,503  & 72,109  & 87,232  & 130,818  & 255,286  \\
 &  &     $d=\infty$ & 7,816  & 8,453  & 9,018  & 9,783  & 37,017  & 39,037  & 50,522  & 78,914  & 72,207  & 70,373  & 90,357  & 199,594  \\
& ONEAudit & $\eta=$0.51 $d=$10 & 8,230  & 8,639  & 9,018  & 9,775  & 46,383  & 49,762  & 59,993  & 82,757  & 84,117  & 95,717  & 138,469  & 259,705  \\
&  & $d=$100 & 8,135  & 8,622  & 9,042  & 9,780  & 43,944  & 48,862  & 59,775  & 82,753  & 77,916  & 91,757  & 135,334  & 258,598  \\
&  & $d=$1000 & 8,116  & 8,701  & 9,088  & 9,789  & 42,090  & 47,062  & 59,549  & 82,893  & 72,035  & 87,880  & 132,796  & 258,431  \\
 &  & $d=\infty$ & 8,189  & 8,732  & 9,208  & 9,805  & 37,260  & 43,588  & 59,033  & 83,958  & 56,357  & 71,377  & 119,701  & 259,498  \\
\cline{2-15} & apKelly & $\eta=$0.52 & 7,962  & 8,469  & 8,708  & 9,527  & 62,935  & 64,271  & 66,275  & 71,029  & 303,860  & 313,931  & 311,775  & 309,128  \\
& ALPHA & $\eta=$0.52 $d=$10 & 8,225  & 8,635  & 9,001  & 9,763  & 46,732  & 49,794  & 59,765  & 82,456  & 84,869  & 96,070  & 137,502  & 257,670  \\
&  &  $d=$100 & 8,151  & 8,589  & 9,012  & 9,768  & 44,281  & 48,886  & 59,471  & 82,439  & 79,160  & 92,479  & 134,418  & 256,283  \\
&  &  $d=$1000 & 7,918  & 8,474  & 8,965  & 9,774  & 42,190  & 46,692  & 58,267  & 82,125  & 74,537  & 87,468  & 129,434  & 253,540  \\
 &  &     $d=\infty$ & \bf{7,771}  & \bf{8,202}  & 8,679  & 9,732  & 63,051  & 53,088  & \bf{48,048}  & 69,938  & 297,773  & 226,001  & 113,478  & 150,425  \\
& ONEAudit & $\eta=$0.52 $d=$10 & 8,225  & 8,640  & 9,024  & 9,778  & 46,312  & 49,832  & 60,321  & 82,961  & 83,836  & 96,004  & 139,309  & 261,953  \\
&  & $d=$100 & 8,145  & 8,606  & 9,049  & 9,783  & 43,979  & 48,885  & 59,968  & 82,937  & 78,223  & 92,171  & 136,033  & 260,821  \\
&  & $d=$1000 & 8,047  & 8,641  & 9,074  & 9,790  & 42,088  & 47,016  & 59,510  & 82,987  & 72,111  & 87,778  & 132,551  & 259,460  \\
 &  & $d=\infty$ & 7,839  & 8,471  & 9,044  & 9,791  & \bf{36,772}  & \bf{39,113}  & 50,998  & 79,471  & 70,015  & 69,134  & \bf{91,287}  & 203,227  \\
\cline{2-15} & apKelly & $\eta=$0.55 & 8,936  & 9,118  & 9,092  & 9,266  & 91,420  & 90,332  & 90,539  & 90,201  & 458,352  & 456,098  & 446,626  & 453,647  \\
& ALPHA & $\eta=$0.55 $d=$10 & 8,217  & 8,638  & 8,994  & 9,762  & 47,181  & 50,008  & 59,613  & 82,380  & 85,866  & 96,812  & 137,675  & 257,513  \\
&  &  $d=$100 & 8,124  & 8,553  & 8,971  & 9,764  & 45,535  & 49,303  & 59,152  & 82,256  & 84,063  & 95,514  & 134,842  & 255,529  \\
&  &  $d=$1000 & 8,140  & 8,433  & 8,791  & 9,738  & 51,308  & 50,587  & 57,507  & 81,018  & 104,307  & 103,285  & 130,298  & 248,422  \\
 &  &     $d=\infty$ & 8,855  & 8,822  & 8,493  & 9,487  & 89,905  & 86,699  & 78,631  & 62,335  & 450,623  & 438,888  & 381,743  & 182,325  \\
& ONEAudit & $\eta=$0.55 $d=$10 & 8,246  & 8,638  & 9,074  & 9,788  & 46,245  & 49,935  & 61,020  & 83,616  & 83,274  & 96,581  & 141,269  & 267,243  \\
&  & $d=$100 & 8,161  & 8,619  & 9,074  & 9,793  & 44,225  & 49,076  & 60,605  & 83,593  & 78,916  & 93,812  & 139,010  & 266,118  \\
&  & $d=$1000 & 7,919  & 8,476  & 9,003  & 9,792  & 42,735  & 47,184  & 59,183  & 83,272  & 76,051  & 88,879  & 133,783  & 262,572  \\
 &  & $d=\infty$ & 7,919  & 8,209  & 8,624  & 9,735  & 70,910  & 60,933  & 49,689  & 68,352  & 349,194  & 293,641  & 136,117  & \bf{146,015}  \\
\cline{2-15} & apKelly & $\eta=$0.6 & 9,289  & 9,269  & 9,288  & 9,430  & 92,975  & 92,124  & 92,959  & 93,401  & 467,898  & 465,514  & 463,709  & 469,873  \\
& ALPHA & $\eta=$0.6 $d=$10 & 8,289  & 8,670  & 8,971  & 9,750  & 48,013  & 50,519  & 59,460  & 82,308  & 89,010  & 98,375  & 138,201  & 256,774  \\
&  &  $d=$100 & 8,364  & 8,605  & 8,942  & 9,755  & 50,576  & 51,474  & 59,582  & 81,981  & 98,741  & 103,538  & 137,676  & 254,412  \\
&  &  $d=$1000 & 8,986  & 8,848  & 8,757  & 9,654  & 72,492  & 65,538  & 61,183  & 78,971  & 199,481  & 167,721  & 150,812  & 243,175  \\
 &  &     $d=\infty$ & 9,223  & 9,163  & 9,077  & \bf{9,200}  & 93,077  & 92,226  & 89,731  & 77,554  & 465,050  & 458,055  & 442,870  & 380,694  \\
& ONEAudit & $\eta=$0.6 $d=$10 & 8,256  & 8,710  & 9,140  & 9,814  & 46,198  & 50,470  & 62,126  & 84,771  & 84,058  & 98,076  & 145,759  & 276,759  \\
&  & $d=$100 & 8,183  & 8,637  & 9,113  & 9,815  & 45,145  & 49,834  & 62,046  & 84,620  & 82,234  & 97,068  & 143,571  & 275,290  \\
&  & $d=$1000 & 8,063  & 8,423  & 8,917  & 9,796  & 48,570  & 50,119  & 59,544  & 83,731  & 94,875  & 101,193  & 137,250  & 268,303  \\
 &  & $d=\infty$ & 8,755  & 8,691  & \bf{8,430}  & 9,614  & 88,367  & 85,547  & 71,029  & \bf{62,327}  & 444,414  & 429,114  & 346,251  & 158,316  
\end{tabular} \caption{\protect \label{tab:performance-detail-1}
Mean sample sizes to reject the null that the assorter mean does not exceed 1/2 when the fraction of valid votes for the winner is 0.505, 
for various population sample sizes, numbers of blank/invalid cards, based on 1,000 replications.
The smallest sample size for each of the 12 conditions is in bold font.
Some flavor of ALPHA applied to the ONEAudit transformation of the assorter values
had the smallest sample size in 6 of the 12 conditions; some flavor of ALPHA applied to the raw assorter values had the smallest in 4 conditions, 
and some flavor of \emph{a priori} Kelly applied to the raw assorter values had the smallest in 2 conditions.}
\end{table}

\begin{table}[ht]
\tiny
\begin{tabular}{llr|rrrr|rrrr|rrrr} 
& & & \multicolumn{4}{|c|}{$N=$10,000, \%blank} &  \multicolumn{4}{|c|}{$N=$100,000 \%blank} & \multicolumn{4}{|c}{$N=$500,000 \%blank} \\ 
$\theta$ & Method & params & 10 & 25 & 50 & 75  & 10 & 25 & 50 & 75  & 10 & 25 & 50 & 75  \\
\hline 0.51 & sqKelly & & 7,394  & 7,416  & 7,468  & 8,100  & 71,259  & 70,984  & 73,780  & 71,692  & 362,456  & 360,640  & 363,556  & 359,819  \\
\cline{2-15} & apKelly & $\eta=$0.505 & 8,295  & 8,759  & 9,490  & 9,946  & 18,821  & 22,636  & 30,713  & 50,214  & 21,200  & 25,247  & 37,491  & 71,246  \\
& ALPHA & $\eta=$0.505 $d=$10 & 6,305  & 6,655  & 7,597  & 9,101  & 19,189  & 22,492  & 31,511  & 56,797  & 23,919  & 28,774  & 46,053  & 113,995  \\
&  &  $d=$100 & 6,104  & 6,601  & 7,600  & 9,128  & 17,462  & 21,409  & 31,102  & 56,897  & 21,477  & 26,774  & 44,840  & 113,564  \\
&  &  $d=$1000 & 6,096  & 6,617  & 7,687  & 9,183  & 16,575  & 20,806  & 31,092  & 57,477  & 20,182  & 25,462  & 44,406  & 114,501  \\
 &  &     $d=\infty$ & 6,329  & 6,901  & 8,024  & 9,296  & 18,036  & 23,373  & 36,453  & 64,506  & 22,101  & 29,744  & 57,651  & 153,556  \\
& ONEAudit & $\eta=$0.505 $d=$10 & 6,301  & 6,658  & 7,604  & 9,110  & 19,127  & 22,514  & 31,591  & 56,982  & 23,813  & 28,768  & 46,223  & 114,612  \\
&  & $d=$100 & 6,104  & 6,618  & 7,611  & 9,138  & 17,496  & 21,444  & 31,212  & 57,127  & 21,473  & 26,805  & 45,047  & 114,337  \\
&  & $d=$1000 & 6,160  & 6,673  & 7,742  & 9,199  & 16,855  & 21,199  & 31,472  & 57,727  & 20,431  & 26,028  & 45,217  & 115,831  \\
 &  & $d=\infty$ & 6,694  & 7,176  & 8,202  & 9,336  & 24,101  & 29,993  & 43,326  & 68,267  & 35,297  & 47,303  & 85,375  & 191,268  \\
\cline{2-15} & apKelly & $\eta=$0.51 & 6,535  & 7,032  & 8,099  & 9,491  & 13,455  & \bf{16,424}  & 21,998  & 36,589  & 15,353  & \bf{18,538}  & 27,698  & 51,396  \\
& ALPHA & $\eta=$0.51 $d=$10 & 6,311  & 6,657  & 7,595  & 9,101  & 19,157  & 22,484  & 31,518  & 56,779  & 23,953  & 28,780  & 46,040  & 113,940  \\
&  &  $d=$100 & 6,095  & 6,568  & 7,585  & 9,125  & 17,545  & 21,265  & 31,055  & 56,806  & 21,537  & 26,764  & 44,821  & 113,369  \\
&  &  $d=$1000 & 5,969  & 6,517  & 7,623  & 9,167  & 16,212  & 20,263  & 30,581  & 57,146  & 19,659  & 24,987  & 43,438  & 113,584  \\
 &  &     $d=\infty$ & 5,812  & 6,405  & 7,702  & 9,225  & \bf{13,587}  & 17,433  & 28,011  & 57,190  & 15,657  & 19,894  & 36,657  & 107,352  \\
& ONEAudit & $\eta=$0.51 $d=$10 & 6,292  & 6,661  & 7,611  & 9,117  & 19,068  & 22,503  & 31,672  & 57,148  & 23,768  & 28,795  & 46,429  & 115,157  \\
&  & $d=$100 & 6,107  & 6,622  & 7,616  & 9,143  & 17,438  & 21,462  & 31,282  & 57,261  & 21,518  & 26,801  & 45,205  & 114,937  \\
&  & $d=$1000 & 6,111  & 6,636  & 7,716  & 9,198  & 16,667  & 20,960  & 31,289  & 57,766  & 20,273  & 25,678  & 44,970  & 115,955  \\
 &  & $d=\infty$ & 6,349  & 6,921  & 8,048  & 9,308  & 18,183  & 23,536  & 36,810  & 64,825  & 22,265  & 30,008  & 58,317  & 155,129  \\
\cline{2-15} & apKelly & $\eta=$0.52 & 5,657  & 5,948  & 6,836  & 8,433  & 17,817  & 20,617  & 24,937  & 35,984  & 34,031  & 39,193  & 50,230  & 74,171  \\
& ALPHA & $\eta=$0.52 $d=$10 & 6,315  & 6,664  & 7,594  & 9,098  & 19,220  & 22,504  & 31,524  & 56,747  & 24,014  & 28,837  & 46,045  & 113,849  \\
&  &  $d=$100 & 6,120  & 6,555  & 7,564  & 9,114  & 17,616  & 21,224  & 30,968  & 56,693  & 21,744  & 26,877  & 44,688  & 113,084  \\
&  &  $d=$1000 & 5,853  & 6,330  & 7,458  & 9,134  & 15,812  & 19,438  & 29,469  & 56,485  & 19,189  & 24,381  & 41,799  & 111,544  \\
 &  &     $d=\infty$ & \bf{5,495}  & 5,815  & 7,116  & 9,054  & 16,431  & 16,861  & 21,695  & 45,940  & 25,327  & 22,107  & 27,601  & 69,750  \\
& ONEAudit & $\eta=$0.52 $d=$10 & 6,288  & 6,671  & 7,628  & 9,129  & 19,029  & 22,509  & 31,793  & 57,565  & 23,725  & 28,774  & 46,722  & 116,513  \\
&  & $d=$100 & 6,098  & 6,609  & 7,632  & 9,156  & 17,492  & 21,475  & 31,424  & 57,549  & 21,650  & 26,912  & 45,512  & 116,168  \\
&  & $d=$1000 & 6,009  & 6,557  & 7,666  & 9,196  & 16,329  & 20,490  & 31,046  & 57,826  & 19,799  & 25,256  & 44,413  & 116,211  \\
 &  & $d=\infty$ & 5,840  & 6,445  & 7,749  & 9,249  & 13,675  & 17,596  & 28,400  & 57,957  & \bf{15,735}  & 20,029  & 37,279  & 109,939  \\
\cline{2-15} & apKelly & $\eta=$0.55 & 7,586  & 7,636  & 7,624  & 8,143  & 73,964  & 74,569  & 77,857  & 75,379  & 379,654  & 379,856  & 385,180  & 376,644  \\
& ALPHA & $\eta=$0.55 $d=$10 & 6,332  & 6,674  & 7,578  & 9,096  & 19,361  & 22,590  & 31,486  & 56,635  & 24,427  & 29,020  & 46,145  & 113,420  \\
&  &  $d=$100 & 6,155  & 6,523  & 7,501  & 9,097  & 18,335  & 21,520  & 30,659  & 56,336  & 23,038  & 27,633  & 44,458  & 112,383  \\
&  &  $d=$1000 & 6,139  & 6,228  & 7,138  & 9,011  & 19,584  & 20,711  & 27,763  & 54,375  & 25,583  & 27,311  & 40,208  & 105,619  \\
 &  &     $d=\infty$ & 7,543  & 6,890  & \bf{6,511}  & 8,506  & 74,146  & 63,101  & 33,353  & 34,003  & 374,407  & 322,840  & 115,810  & 51,300  \\
& ONEAudit & $\eta=$0.55 $d=$10 & 6,285  & 6,702  & 7,700  & 9,176  & 18,936  & 22,696  & 32,228  & 58,523  & 23,654  & 28,942  & 47,848  & 120,450  \\
&  & $d=$100 & 6,122  & 6,620  & 7,667  & 9,194  & 17,700  & 21,680  & 31,893  & 58,562  & 21,877  & 27,271  & 46,624  & 119,742  \\
&  & $d=$1000 & 5,848  & 6,338  & 7,519  & 9,194  & 15,983  & 19,648  & 30,124  & 57,967  & 19,540  & 24,758  & 43,331  & 117,202  \\
 &  & $d=\infty$ & 5,557  & \bf{5,738}  & 7,016  & 9,052  & 20,885  & 18,719  & \bf{21,218}  & 43,862  & 49,275  & 29,202  & \bf{27,591}  & 64,394  \\
\cline{2-15} & apKelly & $\eta=$0.6 & 9,027  & 8,968  & 9,077  & 9,067  & 89,418  & 90,560  & 90,425  & 90,184  & 443,859  & 450,170  & 456,126  & 455,792  \\
& ALPHA & $\eta=$0.6 $d=$10 & 6,395  & 6,666  & 7,583  & 9,090  & 19,737  & 22,988  & 31,536  & 56,452  & 25,199  & 29,536  & 46,230  & 113,089  \\
&  &  $d=$100 & 6,494  & 6,626  & 7,471  & 9,059  & 21,103  & 22,978  & 30,620  & 55,890  & 27,423  & 30,148  & 45,321  & 111,313  \\
&  &  $d=$1000 & 7,848  & 7,233  & 7,014  & 8,813  & 39,196  & 32,645  & 30,132  & 51,644  & 63,444  & 49,191  & 45,960  & 98,789  \\
 &  &     $d=\infty$ & 8,965  & 8,703  & 7,656  & \bf{7,904}  & 88,829  & 87,058  & 79,399  & 41,150  & 438,065  & 430,248  & 392,792  & 124,134  \\
& ONEAudit & $\eta=$0.6 $d=$10 & 6,326  & 6,785  & 7,790  & 9,252  & 18,926  & 23,047  & 33,272  & 60,344  & 23,681  & 29,491  & 49,670  & 127,072  \\
&  & $d=$100 & 6,200  & 6,657  & 7,720  & 9,249  & 18,237  & 22,179  & 32,733  & 60,210  & 22,845  & 28,311  & 48,553  & 126,355  \\
&  & $d=$1000 & 6,002  & 6,201  & 7,355  & 9,193  & 18,238  & 20,520  & 29,486  & 58,343  & 23,405  & 26,534  & 43,312  & 119,646  \\
 &  & $d=\infty$ & 7,049  & 6,486  & 6,539  & 8,723  & 68,264  & 52,660  & 27,505  & \bf{35,331}  & 341,699  & 270,043  & 66,439  & \bf{50,969}  
\end{tabular} \caption{\protect \label{tab:performance-detail-2}
Same as table~\ref{tab:performance-detail-1}, but with a fraction 0.51 of the valid votes for the reported winner.
Some flavor of ALPHA applied to the ONEAudit transformation of the assorter values
had the smallest sample size for 6 conditions; some flavor of ALPHA applied to the raw assorter values had the smallest for
4 conditions, and some flavor of \emph{a priori} Kelly applied to the raw assorter values had the smallest for 2 conditions.
}
\end{table}

\begin{table}[ht]
\tiny
\begin{tabular}{llr|rrrr|rrrr|rrrr} 
& & & \multicolumn{4}{|c|}{$N=$10,000, \%blank} &  \multicolumn{4}{|c|}{$N=$100,000 \%blank} & \multicolumn{4}{|c}{$N=$500,000 \%blank} \\ 
$\theta$ & Method & params & 10 & 25 & 50 & 75  & 10 & 25 & 50 & 75  & 10 & 25 & 50 & 75  \\
\hline 0.52 & sqKelly & & 3,416  & 3,648  & 4,134  & 5,502  & 13,469  & 14,905  & 16,084  & 22,542  & 38,589  & 43,864  & 49,984  & 64,106  \\
\cline{2-15} & apKelly & $\eta=$0.505 & 5,909  & 6,588  & 7,849  & 9,560  & 8,974  & 10,737  & 15,375  & 28,590  & 9,379  & 11,108  & 16,791  & 33,517  \\
& ALPHA & $\eta=$0.505 $d=$10 & 3,454  & 3,798  & 4,989  & 7,544  & 5,637  & 6,941  & 10,800  & 27,035  & 6,089  & 7,304  & 12,313  & 36,836  \\
&  &  $d=$100 & 3,292  & 3,700  & 4,984  & 7,571  & 5,257  & 6,667  & 10,687  & 27,128  & 5,608  & 6,960  & 12,179  & 36,853  \\
&  &  $d=$1000 & 3,361  & 3,880  & 5,239  & 7,732  & 5,333  & 6,913  & 11,268  & 28,144  & 5,717  & 7,216  & 12,865  & 38,268  \\
 &  &     $d=\infty$ & 4,064  & 4,663  & 6,066  & 8,173  & 8,547  & 11,411  & 19,692  & 42,424  & 9,937  & 13,662  & 28,041  & 83,317  \\
& ONEAudit & $\eta=$0.505 $d=$10 & 3,449  & 3,799  & 5,001  & 7,554  & 5,628  & 6,947  & 10,839  & 27,193  & 6,099  & 7,315  & 12,355  & 37,080  \\
&  & $d=$100 & 3,304  & 3,724  & 5,005  & 7,588  & 5,259  & 6,700  & 10,766  & 27,270  & 5,637  & 6,994  & 12,240  & 37,131  \\
&  & $d=$1000 & 3,447  & 3,954  & 5,316  & 7,765  & 5,527  & 7,113  & 11,524  & 28,473  & 5,939  & 7,432  & 13,169  & 38,931  \\
 &  & $d=\infty$ & 4,468  & 5,044  & 6,319  & 8,258  & 12,336  & 15,708  & 24,738  & 46,521  & 16,996  & 23,151  & 43,940  & 108,570  \\
\cline{2-15} & apKelly & $\eta=$0.51 & 3,977  & 4,469  & 5,716  & 7,999  & 5,305  & 6,516  & 9,182  & 17,430  & 5,582  & 6,600  & 9,877  & 19,842  \\
& ALPHA & $\eta=$0.51 $d=$10 & 3,451  & 3,795  & 4,987  & 7,543  & 5,646  & 6,936  & 10,799  & 27,032  & 6,107  & 7,298  & 12,293  & 36,817  \\
&  &  $d=$100 & 3,274  & 3,684  & 4,973  & 7,564  & 5,219  & 6,626  & 10,652  & 27,067  & 5,585  & 6,939  & 12,112  & 36,772  \\
&  &  $d=$1000 & 3,201  & 3,746  & 5,115  & 7,693  & 5,014  & 6,527  & 10,902  & 27,729  & 5,405  & 6,848  & 12,423  & 37,669  \\
 &  &     $d=\infty$ & 3,447  & 4,056  & 5,606  & 8,006  & 5,498  & 7,526  & 13,818  & 35,277  & 5,966  & 7,970  & 16,362  & 54,916  \\
& ONEAudit & $\eta=$0.51 $d=$10 & 3,447  & 3,802  & 5,008  & 7,567  & 5,627  & 6,958  & 10,862  & 27,315  & 6,088  & 7,319  & 12,393  & 37,273  \\
&  & $d=$100 & 3,299  & 3,723  & 5,010  & 7,599  & 5,261  & 6,691  & 10,773  & 27,381  & 5,636  & 6,988  & 12,279  & 37,323  \\
&  & $d=$1000 & 3,386  & 3,900  & 5,279  & 7,761  & 5,372  & 6,957  & 11,390  & 28,430  & 5,776  & 7,274  & 12,988  & 38,855  \\
 &  & $d=\infty$ & 4,081  & 4,687  & 6,093  & 8,195  & 8,631  & 11,513  & 19,900  & 42,767  & 10,022  & 13,816  & 28,401  & 84,337  \\
\cline{2-15} & apKelly & $\eta=$0.52 & 2,833  & 3,206  & 4,158  & 6,151  & \bf{3,985}  & \bf{4,835}  & 6,686  & 12,577  & \bf{4,050}  & \bf{4,968}  & 7,159  & \bf{15,237}  \\
& ALPHA & $\eta=$0.52 $d=$10 & 3,451  & 3,798  & 4,985  & 7,537  & 5,652  & 6,937  & 10,786  & 27,011  & 6,126  & 7,299  & 12,277  & 36,795  \\
&  &  $d=$100 & 3,240  & 3,649  & 4,929  & 7,550  & 5,219  & 6,553  & 10,562  & 26,959  & 5,546  & 6,878  & 11,996  & 36,562  \\
&  &  $d=$1000 & 2,992  & 3,491  & 4,903  & 7,604  & 4,632  & 6,024  & 10,179  & 26,983  & 4,842  & 6,243  & 11,568  & 36,439  \\
 &  &     $d=\infty$ & 2,781  & 3,321  & 4,805  & 7,668  & 4,017  & 5,166  & 8,923  & 25,652  & 4,148  & 5,300  & 9,820  & 32,686  \\
& ONEAudit & $\eta=$0.52 $d=$10 & 3,440  & 3,821  & 5,027  & 7,590  & 5,616  & 6,961  & 10,952  & 27,542  & 6,092  & 7,343  & 12,478  & 37,742  \\
&  & $d=$100 & 3,292  & 3,713  & 5,024  & 7,621  & 5,266  & 6,682  & 10,824  & 27,644  & 5,633  & 7,010  & 12,358  & 37,724  \\
&  & $d=$1000 & 3,242  & 3,790  & 5,184  & 7,746  & 5,080  & 6,660  & 11,125  & 28,349  & 5,466  & 6,962  & 12,686  & 38,721  \\
 &  & $d=\infty$ & 3,479  & 4,110  & 5,663  & 8,053  & 5,601  & 7,651  & 14,122  & 35,919  & 6,046  & 8,120  & 16,771  & 56,462  \\
\cline{2-15} & apKelly & $\eta=$0.55 & 3,592  & 3,748  & 4,237  & \bf{5,423}  & 19,030  & 20,062  & 19,874  & 25,982  & 75,673  & 80,329  & 85,559  & 94,161  \\
& ALPHA & $\eta=$0.55 $d=$10 & 3,465  & 3,804  & 4,967  & 7,526  & 5,713  & 6,951  & 10,746  & 26,930  & 6,180  & 7,322  & 12,252  & 36,697  \\
&  &  $d=$100 & 3,214  & 3,596  & 4,831  & 7,504  & 5,319  & 6,495  & 10,356  & 26,669  & 5,538  & 6,829  & 11,753  & 36,083  \\
&  &  $d=$1000 & 2,950  & 3,200  & 4,370  & 7,349  & 4,901  & 5,628  & 8,834  & 24,842  & 5,168  & 5,870  & 9,762  & 32,989  \\
 &  &     $d=\infty$ & 3,477  & 3,217  & \bf{3,740}  & 6,601  & 14,747  & 8,138  & 6,785  & 14,998  & 44,666  & 12,870  & 7,540  & 17,473  \\
& ONEAudit & $\eta=$0.55 $d=$10 & 3,437  & 3,843  & 5,110  & 7,674  & 5,639  & 7,067  & 11,206  & 28,408  & 6,094  & 7,424  & 12,818  & 39,168  \\
&  & $d=$100 & 3,258  & 3,710  & 5,066  & 7,689  & 5,282  & 6,695  & 10,981  & 28,362  & 5,616  & 7,087  & 12,557  & 39,012  \\
&  & $d=$1000 & 2,973  & 3,486  & 4,951  & 7,714  & 4,604  & 6,030  & 10,375  & 28,075  & 4,836  & 6,242  & 11,796  & 38,400  \\
 &  & $d=\infty$ & \bf{2,701}  & 3,167  & 4,653  & 7,643  & 4,011  & 4,903  & 8,214  & 23,898  & 4,069  & 5,041  & 8,935  & 29,407  \\
\cline{2-15} & apKelly & $\eta=$0.6 & 7,501  & 7,544  & 7,447  & 7,605  & 76,103  & 75,259  & 75,515  & 77,007  & 386,240  & 385,732  & 403,069  & 385,244  \\
& ALPHA & $\eta=$0.6 $d=$10 & 3,527  & 3,822  & 4,926  & 7,500  & 5,883  & 7,052  & 10,732  & 26,811  & 6,347  & 7,423  & 12,165  & 36,439  \\
&  &  $d=$100 & 3,445  & 3,653  & 4,714  & 7,435  & 5,922  & 6,710  & 10,160  & 26,241  & 6,279  & 7,150  & 11,487  & 35,380  \\
&  &  $d=$1000 & 4,591  & 3,961  & 4,073  & 6,886  & 10,615  & 8,421  & 8,536  & 21,880  & 11,811  & 9,324  & 9,684  & 28,900  \\
 &  &     $d=\infty$ & 7,466  & 6,465  & 4,335  & 5,508  & 76,870  & 64,404  & 21,902  & \bf{12,308}  & 356,324  & 325,161  & 100,355  & 15,780  \\
& ONEAudit & $\eta=$0.6 $d=$10 & 3,461  & 3,916  & 5,223  & 7,821  & 5,710  & 7,226  & 11,679  & 29,901  & 6,158  & 7,584  & 13,382  & 41,789  \\
&  & $d=$100 & 3,288  & 3,732  & 5,126  & 7,813  & 5,376  & 6,774  & 11,308  & 29,686  & 5,639  & 7,246  & 12,978  & 41,427  \\
&  & $d=$1000 & 2,900  & 3,271  & 4,663  & 7,667  & 4,702  & 5,725  & 9,589  & 27,903  & 4,919  & 5,948  & 10,705  & 38,147  \\
 &  & $d=\infty$ & 3,132  & \bf{3,041}  & 3,873  & 6,981  & 9,126  & 6,374  & \bf{6,674}  & 16,535  & 17,623  & 8,035  & \bf{7,137}  & 19,129  
\end{tabular} \caption{\protect \label{tab:performance-detail-3}
Same as table~\ref{tab:performance-detail-1}, but with a fraction 0.52 of the valid votes for the reported winner.
Some flavor of ALPHA applied to the ONEAudit transformation of the assorter values had the smallest sample size for 4 conditions; some flavor of ALPHA applied to the raw assorter values had the smallest for
2 conditions, and some flavor of \emph{a priori} Kelly applied to the raw assorter values 
had the smallest for 6 conditions, of which 5 used the true population mean.}
\end{table}

\begin{table}[ht]
\tiny
\begin{tabular}{llr|rrrr|rrrr|rrrr} 
& & & \multicolumn{4}{|c|}{$N=$10,000, \%blank} &  \multicolumn{4}{|c|}{$N=$100,000 \%blank} & \multicolumn{4}{|c}{$N=$500,000 \%blank} \\ 
$\theta$ & Method & params & 10 & 25 & 50 & 75  & 10 & 25 & 50 & 75  & 10 & 25 & 50 & 75  \\
\hline 0.55 & sqKelly & & 594  & 748  & 1,067  & 1,834  & 688  & \em{845}  & 1,208  & \em{2,331}  & 705  & 787  & 1,213  & 2,446  \\
\cline{2-15} & apKelly & $\eta=$0.505 & 2,919  & 3,426  & 4,634  & 7,180  & 3,435  & 4,144  & 6,098  & 11,800  & 3,526  & 4,167  & 6,242  & 12,311  \\
& ALPHA & $\eta=$0.505 $d=$10 & 790  & 1,009  & 1,625  & 3,577  & 946  & 1,180  & 1,973  & 5,799  & 939  & 1,113  & 2,044  & 6,158  \\
&  &  $d=$100 & 790  & 1,043  & 1,694  & 3,683  & 930  & 1,187  & 2,062  & 6,012  & 935  & 1,155  & 2,117  & 6,360  \\
&  &  $d=$1000 & 1,057  & 1,342  & 2,114  & 4,154  & 1,234  & 1,582  & 2,690  & 7,135  & 1,245  & 1,577  & 2,773  & 7,536  \\
 &  &     $d=\infty$ & 1,800  & 2,217  & 3,315  & 5,500  & 3,284  & 4,501  & 8,283  & 20,041  & 3,767  & 5,221  & 10,930  & 34,385  \\
& ONEAudit & $\eta=$0.505 $d=$10 & 791  & 1,012  & 1,633  & 3,593  & 945  & 1,185  & 1,984  & 5,842  & 941  & 1,114  & 2,056  & 6,195  \\
&  & $d=$100 & 799  & 1,054  & 1,710  & 3,706  & 942  & 1,202  & 2,083  & 6,067  & 945  & 1,170  & 2,138  & 6,417  \\
&  & $d=$1000 & 1,108  & 1,395  & 2,174  & 4,205  & 1,308  & 1,662  & 2,792  & 7,276  & 1,317  & 1,656  & 2,869  & 7,693  \\
 &  & $d=\infty$ & 2,084  & 2,489  & 3,550  & 5,648  & 4,965  & 6,489  & 10,806  & 22,868  & 6,750  & 9,133  & 17,700  & 46,539  \\
\cline{2-15} & apKelly & $\eta=$0.51 & 1,659  & 1,969  & 2,809  & 4,689  & 1,831  & 2,220  & 3,256  & 6,430  & 1,856  & 2,203  & 3,322  & 6,561  \\
& ALPHA & $\eta=$0.51 $d=$10 & 790  & 1,006  & 1,623  & 3,574  & 945  & 1,179  & 1,972  & 5,796  & 936  & 1,110  & 2,038  & 6,154  \\
&  &  $d=$100 & 778  & 1,024  & 1,675  & 3,668  & 916  & 1,173  & 2,043  & 5,982  & 921  & 1,135  & 2,092  & 6,339  \\
&  &  $d=$1000 & 968  & 1,254  & 2,020  & 4,084  & 1,122  & 1,459  & 2,542  & 6,955  & 1,132  & 1,448  & 2,610  & 7,346  \\
 &  &     $d=\infty$ & 1,395  & 1,799  & 2,906  & 5,236  & 1,928  & 2,710  & 5,431  & 15,712  & 2,023  & 2,826  & 6,067  & 21,572  \\
& ONEAudit & $\eta=$0.51 $d=$10 & 792  & 1,011  & 1,637  & 3,611  & 944  & 1,186  & 1,992  & 5,875  & 941  & 1,116  & 2,064  & 6,236  \\
&  & $d=$100 & 796  & 1,052  & 1,710  & 3,714  & 938  & 1,199  & 2,083  & 6,102  & 941  & 1,166  & 2,140  & 6,452  \\
&  & $d=$1000 & 1,068  & 1,352  & 2,137  & 4,186  & 1,248  & 1,601  & 2,726  & 7,224  & 1,256  & 1,593  & 2,801  & 7,641  \\
 &  & $d=\infty$ & 1,816  & 2,236  & 3,339  & 5,531  & 3,320  & 4,549  & 8,378  & 20,256  & 3,806  & 5,285  & 11,077  & 34,892  \\
\cline{2-15} & apKelly & $\eta=$0.52 & 952  & 1,163  & 1,673  & 2,930  & 1,045  & 1,265  & 1,858  & 3,656  & 1,047  & 1,242  & 1,869  & 3,749  \\
& ALPHA & $\eta=$0.52 $d=$10 & 790  & 1,004  & 1,616  & 3,570  & 944  & 1,173  & 1,968  & 5,781  & 932  & 1,108  & 2,028  & 6,143  \\
&  &  $d=$100 & 751  & 997  & 1,646  & 3,638  & 891  & 1,144  & 2,001  & 5,921  & 896  & 1,100  & 2,047  & 6,280  \\
&  &  $d=$1000 & 821  & 1,096  & 1,853  & 3,944  & 955  & 1,250  & 2,274  & 6,580  & 960  & 1,238  & 2,328  & 6,974  \\
 &  &     $d=\infty$ & 948  & 1,283  & 2,266  & 4,721  & 1,117  & 1,548  & 3,160  & 10,491  & 1,126  & 1,562  & 3,304  & 11,965  \\
& ONEAudit & $\eta=$0.52 $d=$10 & 793  & 1,015  & 1,649  & 3,648  & 945  & 1,190  & 2,009  & 5,954  & 942  & 1,120  & 2,077  & 6,322  \\
&  & $d=$100 & 786  & 1,044  & 1,712  & 3,737  & 926  & 1,191  & 2,087  & 6,150  & 930  & 1,157  & 2,144  & 6,507  \\
&  & $d=$1000 & 983  & 1,278  & 2,060  & 4,153  & 1,147  & 1,492  & 2,602  & 7,134  & 1,153  & 1,477  & 2,676  & 7,541  \\
 &  & $d=\infty$ & 1,420  & 1,830  & 2,955  & 5,302  & 1,965  & 2,767  & 5,564  & 16,093  & 2,067  & 2,889  & 6,242  & 22,239  \\
\cline{2-15} & apKelly & $\eta=$0.55 & \bf{581}  & \bf{737}  & 1,054  & \bf{1,816}  & \bf{673}  & \bf{844}  & \bf{1,186}  & \bf{2,330}  & 692  & \bf{780} & \bf{1,189}  & \bf{2,413}  \\
& ALPHA & $\eta=$0.55 $d=$10 & 788  & 996  & 1,606  & 3,553  & 939  & 1,163  & 1,951  & 5,742  & 932  & 1,099  & 2,006  & 6,106  \\
&  &  $d=$100 & 696  & 926  & 1,557  & 3,566  & 826  & 1,066  & 1,887  & 5,749  & 836  & 1,020  & 1,931  & 6,099  \\
&  &  $d=$1000 & 616  & 831  & 1,458  & 3,540  & 715  & 945  & 1,718  & 5,633  & 725  & 891  & 1,750  & 5,912  \\
 &  &     $d=\infty$ & 594  & 790  & 1,381  & 3,480  & 681  & 884  & 1,581  & 5,200  & \bf{691}  & 846  & 1,589  & 5,400  \\
& ONEAudit & $\eta=$0.55 $d=$10 & 795  & 1,029  & 1,683  & 3,739  & 945  & 1,204  & 2,059  & 6,194  & 951  & 1,137  & 2,127  & 6,591  \\
&  & $d=$100 & 764  & 1,023  & 1,714  & 3,803  & 903  & 1,176  & 2,092  & 6,331  & 910  & 1,129  & 2,148  & 6,705  \\
&  & $d=$1000 & 797  & 1,079  & 1,867  & 4,056  & 925  & 1,229  & 2,287  & 6,870  & 928  & 1,215  & 2,344  & 7,275  \\
 &  & $d=\infty$ & 863  & 1,177  & 2,143  & 4,664  & 1,000  & 1,377  & 2,825  & 9,605  & 999  & 1,366  & 2,922  & 10,626  \\
\cline{2-15} & apKelly & $\eta=$0.6 & 1,044  & 1,238  & 1,557  & 2,143  & 2,795  & 3,568  & 3,811  & 5,811  & 6,072  & 5,010  & 7,595  & 11,746  \\
& ALPHA & $\eta=$0.6 $d=$10 & 787  & 995  & 1,585  & 3,527  & 939  & 1,155  & 1,930  & 5,690  & 940  & 1,096  & 1,974  & 6,050  \\
&  &  $d=$100 & 689  & 872  & 1,444  & 3,430  & 812  & 999  & 1,724  & 5,483  & 815  & 949  & 1,783  & 5,805  \\
&  &  $d=$1000 & 685  & 775  & 1,149  & 2,977  & 845  & 908  & 1,334  & 4,395  & 870  & 840  & 1,347  & 4,622  \\
 &  &     $d=\infty$ & 898  & 831  & \bf{1,045}  & 2,446  & 1,537  & 1,112  & 1,187  & 3,097  & 2,107  & 1,011  & \bf{1,189}  & 3,183  \\
& ONEAudit & $\eta=$0.6 $d=$10 & 802  & 1,057  & 1,753  & 3,905  & 951  & 1,231  & 2,153  & 6,596  & 957  & 1,170  & 2,224  & 7,056  \\
&  & $d=$100 & 730  & 1,002  & 1,724  & 3,922  & 868  & 1,157  & 2,108  & 6,629  & 879  & 1,099  & 2,162  & 7,072  \\
&  & $d=$1000 & 646  & 888  & 1,607  & 3,915  & 750  & 1,007  & 1,916  & 6,511  & 752  & 960  & 1,966  & 6,889  \\
 &  & $d=\infty$ & 616  & 835  & 1,514  & 3,851  & 703  & 931  & 1,746  & 6,020  & 706  & 896  & 1,772  & 6,269 
\end{tabular} \caption{\protect \label{tab:performance-detail-4}
Same as table~\ref{tab:performance-detail-1}, but with a fraction 0.55 of the valid votes for the reported winner.
ALPHA applied to the ONEAudit transformation never had the smallest sample size; some flavor of ALPHA applied to
the raw assorter values had the smallest (or was tied for smallest) in
3~conditions, and \emph{a priori} Kelly using the true population mean had the smallest (or was tied for smallest) in 10~conditions.}
\end{table}

\begin{table}[ht]
\tiny
\begin{tabular}{llr|rrrr|rrrr|rrrr} 
& & & \multicolumn{4}{|c|}{$N=$10,000, \%blank} &  \multicolumn{4}{|c|}{$N=$100,000 \%blank} & \multicolumn{4}{|c}{$N=$500,000 \%blank} \\ 
$\theta$ & Method & params & 10 & 25 & 50 & 75  & 10 & 25 & 50 & 75  & 10 & 25 & 50 & 75  \\
\hline 0.6 & sqKelly & & 199  & 247  & 348  & 668  & 210  & 239  & 376  & 729  & 202  & 240  & 368  & 715  \\
\cline{2-15} & apKelly & $\eta=$0.505 & 1,564  & 1,875  & 2,629  & 4,688  & 1,705  & 2,026  & 3,038  & 5,974  & 1,695  & 2,049  & 3,078  & 6,085  \\
& ALPHA & $\eta=$0.505 $d=$10 & 221  & 285  & 478  & 1,324  & 234  & 284  & 522  & 1,597  & 228  & 282  & 519  & 1,560  \\
&  &  $d=$100 & 264  & 336  & 565  & 1,458  & 282  & 342  & 616  & 1,765  & 273  & 342  & 609  & 1,730  \\
&  &  $d=$1000 & 451  & 564  & 897  & 1,985  & 497  & 601  & 1,022  & 2,531  & 492  & 611  & 1,022  & 2,530  \\
 &  &     $d=\infty$ & 924  & 1,167  & 1,819  & 3,459  & 1,635  & 2,211  & 4,199  & 10,631  & 1,812  & 2,578  & 5,469  & 17,416  \\
& ONEAudit & $\eta=$0.505 $d=$10 & 222  & 286  & 481  & 1,333  & 235  & 285  & 525  & 1,610  & 229  & 283  & 523  & 1,570  \\
&  & $d=$100 & 268  & 341  & 573  & 1,472  & 286  & 348  & 625  & 1,785  & 277  & 347  & 618  & 1,750  \\
&  & $d=$1000 & 478  & 592  & 927  & 2,017  & 527  & 634  & 1,062  & 2,589  & 525  & 646  & 1,060  & 2,586  \\
 &  & $d=\infty$ & 1,086  & 1,333  & 1,983  & 3,593  & 2,505  & 3,242  & 5,553  & 12,303  & 3,304  & 4,559  & 8,922  & 23,820  \\
\cline{2-15} & apKelly & $\eta=$0.51 & 832  & 1,000  & 1,448  & 2,672  & 882  & 1,046  & 1,573  & 3,113  & 880  & 1,057  & 1,573  & 3,113  \\
& ALPHA & $\eta=$0.51 $d=$10 & 220  & 284  & 477  & 1,322  & 233  & 283  & 521  & 1,596  & 227  & 280  & 518  & 1,556  \\
&  &  $d=$100 & 257  & 329  & 555  & 1,449  & 275  & 333  & 608  & 1,754  & 265  & 336  & 599  & 1,717  \\
&  &  $d=$1000 & 409  & 519  & 850  & 1,935  & 445  & 549  & 958  & 2,451  & 443  & 556  & 960  & 2,450  \\
 &  &     $d=\infty$ & 697  & 912  & 1,539  & 3,219  & 933  & 1,295  & 2,690  & 8,163  & 960  & 1,372  & 2,992  & 10,794  \\
& ONEAudit & $\eta=$0.51 $d=$10 & 223  & 286  & 483  & 1,341  & 235  & 286  & 527  & 1,622  & 229  & 284  & 524  & 1,580  \\
&  & $d=$100 & 265  & 339  & 571  & 1,477  & 284  & 345  & 624  & 1,789  & 275  & 345  & 616  & 1,755  \\
&  & $d=$1000 & 456  & 569  & 906  & 2,005  & 502  & 607  & 1,033  & 2,565  & 498  & 618  & 1,034  & 2,561  \\
 &  & $d=\infty$ & 932  & 1,178  & 1,835  & 3,486  & 1,651  & 2,235  & 4,250  & 10,751  & 1,829  & 2,610  & 5,545  & 17,659  \\
\cline{2-15} & apKelly & $\eta=$0.52 & 450  & 540  & 791  & 1,511  & 471  & 548  & 835  & 1,667  & 465  & 554  & 834  & 1,643  \\
& ALPHA & $\eta=$0.52 $d=$10 & 218  & 281  & 475  & 1,317  & 232  & 281  & 518  & 1,591  & 226  & 279  & 516  & 1,553  \\
&  &  $d=$100 & 243  & 314  & 539  & 1,429  & 260  & 318  & 589  & 1,729  & 251  & 321  & 581  & 1,695  \\
&  &  $d=$1000 & 341  & 443  & 758  & 1,839  & 367  & 464  & 848  & 2,305  & 361  & 469  & 847  & 2,296  \\
 &  &     $d=\infty$ & 450  & 615  & 1,149  & 2,791  & 512  & 702  & 1,524  & 5,297  & 511  & 722  & 1,562  & 5,917  \\
& ONEAudit & $\eta=$0.52 $d=$10 & 222  & 287  & 486  & 1,356  & 235  & 287  & 531  & 1,641  & 229  & 285  & 528  & 1,602  \\
&  & $d=$100 & 262  & 335  & 570  & 1,485  & 279  & 341  & 621  & 1,804  & 271  & 341  & 614  & 1,769  \\
&  & $d=$1000 & 417  & 529  & 867  & 1,978  & 455  & 560  & 983  & 2,516  & 452  & 568  & 980  & 2,511  \\
 &  & $d=\infty$ & 711  & 932  & 1,570  & 3,278  & 951  & 1,326  & 2,767  & 8,375  & 981  & 1,404  & 3,079  & 11,127  \\
\cline{2-15} & apKelly & $\eta=$0.55 & 221  & 269  & 384  & 737  & 230  & 263  & 409  & 810  & 223  & 264  & 404  & 787  \\
& ALPHA & $\eta=$0.55 $d=$10 & 213  & 276  & 468  & 1,307  & 228  & 277  & 508  & 1,575  & 222  & 272  & 509  & 1,541  \\
&  &  $d=$100 & 210  & 280  & 492  & 1,371  & 223  & 279  & 540  & 1,654  & 215  & 278  & 533  & 1,620  \\
&  &  $d=$1000 & 225  & 304  & 565  & 1,578  & 238  & 307  & 613  & 1,920  & 231  & 306  & 612  & 1,901  \\
 &  &     $d=\infty$ & 233  & 318  & 625  & 1,894  & 246  & 326  & 688  & 2,487  & 239  & 327  & 686  & 2,506  \\
& ONEAudit & $\eta=$0.55 $d=$10 & 223  & 290  & 499  & 1,402  & 236  & 290  & 543  & 1,703  & 231  & 288  & 539  & 1,669  \\
&  & $d=$100 & 247  & 322  & 562  & 1,509  & 264  & 326  & 615  & 1,839  & 255  & 328  & 608  & 1,809  \\
&  & $d=$1000 & 328  & 430  & 760  & 1,896  & 352  & 453  & 848  & 2,389  & 345  & 457  & 849  & 2,381  \\
 &  & $d=\infty$ & 401  & 553  & 1,068  & 2,741  & 441  & 608  & 1,341  & 4,817  & 441  & 619  & 1,362  & 5,233  \\
\cline{2-15} & apKelly & $\eta=$0.6 & \bf{163}  & \bf{204}  & \bf{287}  & \bf{542}  & \bf{173}  & \bf{197}  & \bf{309}  & \bf{627}  & \bf{168}  & \bf{199}  & \bf{307}  & \bf{596}  \\
& ALPHA & $\eta=$0.6 $d=$10 & 208  & 269  & 455  & 1,288  & 224  & 269  & 496  & 1,556  & 218  & 264  & 498  & 1,519  \\
&  &  $d=$100 & 181  & 243  & 427  & 1,279  & 191  & 236  & 470  & 1,540  & 186  & 235  & 465  & 1,503  \\
&  &  $d=$1000 & 167  & 222  & 391  & 1,232  & 176  & 213  & 426  & 1,450  & 171  & 216  & 418  & 1,427  \\
 &  &     $d=\infty$ & 165  & 219  & 380  & 1,200  & 175  & 209  & 408  & 1,364  & 169  & 212  & 404  & 1,361  \\
& ONEAudit & $\eta=$0.6 $d=$10 & 225  & 299  & 521  & 1,484  & 239  & 297  & 569  & 1,821  & 233  & 294  & 567  & 1,788  \\
&  & $d=$100 & 228  & 306  & 552  & 1,559  & 242  & 306  & 606  & 1,911  & 233  & 307  & 598  & 1,883  \\
&  & $d=$1000 & 244  & 331  & 633  & 1,785  & 258  & 340  & 696  & 2,214  & 251  & 341  & 689  & 2,205  \\
 &  & $d=\infty$ & 251  & 348  & 703  & 2,141  & 266  & 360  & 779  & 2,909  & 259  & 360  & 779  & 2,960 
\end{tabular}
 \caption{\protect \label{tab:performance-detail-5}
Same as table~\ref{tab:performance-detail-1}, but with a fraction 0.6 of the valid votes for the reported winner.
\emph{A priori} Kelly applied to the raw assorter values using the correct population mean had the smallest sample 
size for all 12~conditions.}
\end{table}

\end{document}